\documentclass[journal]{IEEEtran}
\ifCLASSINFOpdf
\else
\fi
\usepackage{textcomp,booktabs}
\usepackage[usenames,dvipsnames]{color}
\usepackage{colortbl}
\usepackage{rotating}
\usepackage{amsmath}
\usepackage{amssymb}
\usepackage{graphicx}
\usepackage{booktabs}
\usepackage{tabularx}
\usepackage{array}
\usepackage{lscape}
\usepackage{longtable}
\usepackage{multirow}
\usepackage{epsfig}
\usepackage{epsf}
\usepackage{algorithm}
\usepackage{algorithmic}
\usepackage{graphicx,psfrag,epsfig}
\usepackage[numbers,sort&compress]{natbib}
\newtheorem{theorem}{Lemma}
\ifCLASSOPTIONcompsoc
  \usepackage[caption=false,font=normalsize,labelfont=sf,textfont=sf]{subfig}
\else
  \usepackage[caption=false,font=footnotesize]{subfig}
\fi
\begin{document}
%
\title{Small disturbances can trigger cascading failures in power grids}
%
%
%

\author{Yubo Huang,
        Junguo Lu,
        Weidong Zhang
\thanks{This paper is partly supported by the National Science Foundation of China (61473183，U1509211，61627810), and National Key R\&D Program of China (SQ2017YFGH001005). \emph{Corresponding author: Weidong Zhang}.}
\thanks{The authors are with the Department of Automation, Shanghai Jiao Tong university, Shanghai 200240, China (E-mail: huangyubo@sjtu.edu.cn; jglu@sjtu.edu.cn; wdzhang@sjtu.edu,cn).}}
\maketitle

\begin{abstract}
With the sharp increase of power demand, large-scale blackouts in power grids occur frequently around the world. Cascading failures are the main causes of network outages. Therefore, revealing the complicated cascade mechanism in grids is conducive to design the efficient policy to restrain the failures and further ensure a stable supply of power to users. Motivated by the recent advances of network dynamics, we proposed a framework based Lyapunov stability to analyze the dynamically induced cascading failures in complex networks. We abandoned the assumption that the network is stable in traditional static failure models and then detected that small disturbances actually can trigger cascading failures in unstable networks. What's worse, such failure usually accompanied the overload failure of lines during the equilibriums conversion process. Through the simulation analysis of the Spanish grid, we summarized that the features of this new failure mode include low incidence, large destructiveness, and fast propagation speed. And it usually tends to occur first in edge nodes and subsequently propagate to the central nodes. These features are consistent with the empirical observation of outages in practice.
\end{abstract}


%
\IEEEpeerreviewmaketitle

\section{Introduction}
%
%
%
%

\IEEEPARstart{T}{he} efficient operation of society and individuals is inseparable from a variety of networks. Communication networks promote information flow and cultural exchanges between distant regions~\cite{newman2003structure}; Transportation networks have greatly increased the convenience of travel~\cite{lopez2017spatiotemporal}; Protein networks enable the body to function properly~\cite{xia2019bacteria}; We are so heavily dependent on these natural or man-made networks that we will be troubled when networks crash. The most typical example is the power outage. For instance, the German grid closed a $380,000V$ cable that caused a blackout in Europe in 2006~\cite{ucte2007final}. In 2008, the plant fault caused by snowstorms led to large-scale power outages in Hunan Province, China~\cite{China2008}. In 2009, the storm destroyed several substations and caused 67 million people in Brazil to fall into power outages~\cite{wiki02}. In 2012, the circuit overload during the peak period of power consumption caused a blackout in India~\cite{wiki01}. These blackouts have negatively affected millions of people and caused serious economic loss. From the listed power outages cases, we can straightforwardly conclude that the paralysis of the entire network often triggered by the breakdown of few3 nodes or edges (e.g. the fault of plants or transmission lines caused human or natural factors)~\cite{ma2017fast,ucte2007final,huang2018extension}. Many studies indicate that cascading propagation mechanism is the main culprit in inducing network from local failure to global failure~\cite{jiao2016ds}. What external factors can cause the failure of nodes or lines~\cite{che2018mitigating,che2018fast} and how the failure of one node or line propagates to its neighbors are still two controversial issues in complex networks. Recently, many researchers attempt to answer the two questions from the perspective of dynamics and have made significant advances~\cite{ren2018stochastic,yao2015multi,song2015dynamic}.

In power grids, normally, all nodes (plants) operates synchronically at the equilibrium with the standard frequency $\Omega$~\cite{lozano2012role}. The flow of a transmission line depends on the phase difference between the nodes at both ends. In sync state, the phase of each node is locked so that the grid can generate steady flows to users~\cite{manik2017cycle}. The sync state is highly susceptible to external factors (i.e. attacks~\cite{xiang2018robustness}) that can cause plants fault~\cite{machowski2011power}. Then the network may collapse duo to cascade propagation. Early research on cascading failures mainly focused on network topologies~\cite{dey2016impact} and found that the vulnerability of the network to intentional attacks or random failures depends on the structure of the network~\cite{Albert2004Structural}. For example, the heterogeneous networks
such as scale-free (SF) networks display unexpected robustness to random failures but is vulnerable to intentional attacks. The error and attack tolerance is equivalent to homogeneous networks (e.g. ER networks)~\cite{Albert2004Error}. Therefore, numerous strategies are proposed to resist cascading failure by optimizing the topology of networks~\cite{cao2013improving}. These topology-based static methods indeed obtained some desirable results, yet fail to detect all failures during the power outage since they ignored the exceptional dynamic behaviors of nodes and flows~\cite{schafer2019dynamical,simonsen2008transient,huang2019stability}. To solve this problem, many dynamic models~\cite{motter2019spontaneous,anderson2003power,bergen1981structure,nishikawa2015comparative} are proposed to study the transient dynamics of AC power grids and the most widely accepted is the coarse-scale swing equation (Kuramoto-like model)~\cite{filatrella2008analysis}. In this model, each node is regarded as an oscillator and different nodes are coupled by a sinusoidal function which can spontaneously direct the network to a synchronous state~\cite{rohden2012self}. Therefore, this model can clearly describe the transient dynamics of the entire network and then captures the exceptional dynamic behaivors which may trigger cascading failures.

In this article, we proposed a framework based on network dynamics to reveal the complicate occurrence and propagation mechanisms of cascading failures. First, the swing equation was introduced to describe the coupling dynamics of nodes. Based on the dynamic equation, the Newton downhill algorithm was utilized to solve the equilibrium of the network. Then, we defined the Lyapunov stability of complex networks and derived the criterions of Lyapunov stability to judge the stability of an equilibrium. We found that small disturbances can induce the shutdown of the nodes with exceptional dynamics and further trigger cascade in unstable networks. Through the simulation analysis of the Spanish grid, we concluded that this kind of failure was difficult to occur but could cause power outages throughout the grid once it happens. Faults tended to occur first in less-degree nodes and then indirectly cause the failure of hub nodes through cascade propagation. The propagation speed of the cascade was so quick that it can easily escape the grid protection mechanism, and that implies most control methods are useless for this failure mode. Furthermore, such failure usually accompanied the other failure mode named overload failure of lines during the equilibria conversion process and it accelerates the paralysis of the network. Finally, the dynamic framework discussed above is not limited to power grids and can apply to other networks, such as communication networks, micro-circuit networks, etc., just with appropriate modifications.

\section{The stability analysis of power grids} \label{section1}
In this section, the swing equation is introduced to describe the dynamics of power grids and we subsequently utilize the Newton downhill method to solve the equilibria of grids based on the swing equation. Afterward, the Lyapunov method is used for determining the stability of the obtained equilibria.

\subsection{Modeling power grids}
\begin{figure}
  \centering
  \includegraphics[width=8.5cm]{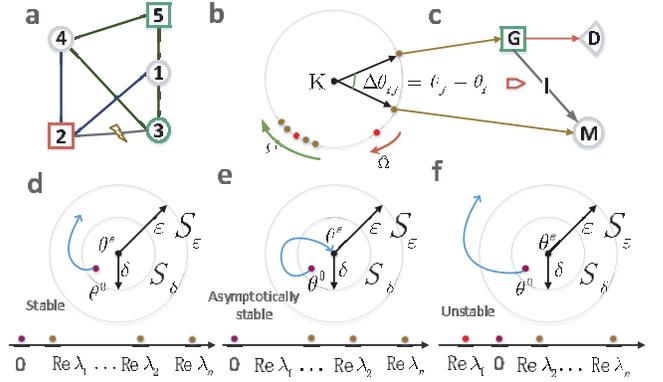}
  \caption{The schematic diagram of a small power grid: \textbf{a} The graph of the small grid with two generators (squares) and three consumers (circles). In our simulation, $M_i = 1$, $D_i = 0.6$, $K = 1.63$ \textbf{b} The green points (rotating machines) run synchronously with frequency $\Omega$, whereas the red points lose synchronization. \textbf{c} The power generated by generator $G$ includes damping dissipation and transmission to consumer $C$. The flow $I$ is determined by the phase difference of node $G$ and node $M$. \textbf{d}-\textbf{f}, $Re(\lambda_1, \lambda_2,\cdots, \lambda_n)$ are the real parts of the eigenvalues of matrix $H$, $\boldsymbol{\theta^0}$ denotes the initial state of the system and $\theta^e$ denotes the equilibrium of the system. $S_{\varepsilon}$ is the restricted area of $f(\boldsymbol{\theta(t)})$ and $S_{\delta}$ is the restricted area of the initial equilibrium $\boldsymbol{\theta^0}$. \textbf{d} is the state track of the stable network. \textbf{e} is the state track of the asymptotically stable network and \textbf{f} is the state track of the unstable network. }\label{smallnetwork}
\end{figure}

A power grid can be abstracted into a weighted graph $G = (V, E)$ (Fig.~\ref{smallnetwork}\textbf{a}), where $V (|V| = N)$ is the vertices set and $E (|E| = L)$ is the edges set of $G$. One vertex can be a generator (generating power) or consumer (consuming power) but they are all regarded as rotating machines (oscillators) in swing equation (Fig.~\ref{smallnetwork}\textbf{b}). If two vertices are linked (through the transmission line), $G(i, j) = 1$, otherwise, $G(i, j) = 0$. The state of oscillators $i$ is completely characterized by its phase $\theta_i$ and the phase velocity $\omega_i = \dot{\theta_i}$ relative to the reference frequency $\Omega = 2\pi = 50/60 Hz$ of the electric system. The flow of $E_{ij}$ depends on the phase difference $\Delta \theta_{ij}$ of $V_i$ and $V_j$ (Fig.~\ref{smallnetwork}\textbf{c}). Thereby, to guarantee steady power flows in grids, all oscillators should run at the same frequency ($\forall i, \omega_i = 0.$ In this state, the phases of all machines are locked), which is called synchronization. How to force all machines to operate synchronously? The answer is the coupling effect between linked oscillators. In the swing equation~\cite{filatrella2008analysis}, the sufficiently large coupling strength $K_{ij} = k_{ij} G_{ij}$, governed by the topology $G_{ij}$ of the network and the interaction strength $k_{ij}$ of oscillators (In power grids, $k_{ij} = B_{ij} U_i U_j$, where $B_{ij}$ is the susceptance between two machines and $U$ is the voltage of the grid), will urge the local group of oscillators to be in step and further direct the entire network to synchronize. $M_i$ is the inertia term of oscillator $i$ which is usually assumed to be $1$. $D_{i}$ is damping constant of the oscillator $i$ and we will prove that it determines the transition time of the oscillator from one state to another in subsection~\ref{second failure}. $P_i$ denotes the power generated (consumed) by the machine including damping and electric power exchanged with its neighbors (Fig.~\ref{smallnetwork}\textbf{c}). Here, we use per unit ($p.u.$) to quantify $P_i$, where $P_{p.u.} = 1 s^{-2} = 100MW$.
\begin{equation}\label{swing eqaution}
\begin{split}
  f(\boldsymbol{\theta}, t) & = M_i \frac{d^2}{dt^2}\theta_i(t) + D_i \frac{d}{dt}\theta_i(t) \\
    & = P_i + \sum_{j = 1}^{N} K_{ij} \sin(\theta_j(t) - \theta_i(t))
\end{split}
\end{equation}
After establishing the dynamic equation of each oscillator, the flow on $E_{ij}$ at time $t$ can be computed by:
\begin{equation}\label{flow}
  F_{ij}(t) = K_{ij} \sin(\theta_j(t) - \theta_i(t))
\end{equation}
From the above analysis, we can summarize the prerequisite for stable operation of the power system is that the phases of all oscillators are locked and it means the system is in sync at the reference $\Omega$. The desired synchronous state of the system is denoted as $\boldsymbol{\theta^e} = [\theta_1^e, \theta_2^e, \cdots, \theta_N^e]$. In control theory, this state is also called the equilibrium duo to $\forall i, \ddot{\boldsymbol{\theta}} = \dot{\boldsymbol{\theta}} = \boldsymbol{0}$. Therefore, the swing equation becomes:
\begin{equation}\label{equilibria}
  \left\{ {\begin{array}{*{20}{c}}
   0 = P_i + \sum_{j = 1}^{N} K_{ij}\sin(\theta_j^e - \theta_i^e)\\
   s.t. \quad \sum_{i = 1}^{N} P_i = 0
  \end{array}} \right.
\end{equation}
For large-scale networks, solving these nonlinear equations presents challenge duo to the strict restriction\footnote{For small network, this equations can be solve by the @fslove tool in MATLAB}. In this paper, we use Newton downhill algorithm (Algorithm~\ref{Newton}) to solve these equations, where $f$ is the Eq.~(\ref{equilibria}), $\varepsilon_1$ and $\varepsilon _2$ are the error coefficients, and $\delta$ is a small random vector used for avoiding the infinite loops.
\begin{algorithm}
\caption{Newton Downhill Algorithm}\label{Newton}
\textbf{Input:} Network $G$, $f$, $\varepsilon_1$, $\varepsilon _2$, $\delta$\\
\textbf{Output:} $\theta^e$
 \begin{algorithmic}[1]
    \FOR {$i = 1$ to $N$}
    \STATE $\theta^0_i = 0$;     \% Initialization; $\theta_i$ can also be randomly generated
    \ENDFOR
    \STATE $\lambda = 1, k = 0$;     \% $\lambda$ is the downhill factor
    \FOR {$i = 1$ to $N$}
    \STATE $\theta^{k+1}_i = \theta^{k}_i - \lambda \frac{f_i(\theta^k)}{f'_i(\theta^k)}$;     \% Newton iteration
    \ENDFOR
    \IF {$|f(\theta^{k+1})| < |f(\theta^{k})|$}{
        \IF {$|\theta^{k+1} - \theta^{k}| < \varepsilon_1$}{
        \STATE $\theta^e = \theta^{k+1}, Return$;      \% satisfy the termination condition
        }
        \ELSE
        \STATE $\theta^k = \theta^{k+1}, \theta^{k+1} = 0, k = k + 1$;     \% Enter the next iteration
        \STATE back to step 5
        \ENDIF
    }
    \ENDIF
    \IF {$|f(\theta^{k+1})| >= |f(\theta^{k})|$}{
         \IF {$\lambda < \varepsilon_{\lambda}$ and $|f(x^{k+1})| <= \varepsilon_2$}
              \STATE $\theta^e = \theta^{k}, Return$;     \% Satisfy the termination condition
          \ELSE
          \IF{$\lambda < \varepsilon_{\lambda}$ and $|f(x^{k+1})| > \varepsilon_2$}{
          \STATE $\theta^{k} = \theta^{k+1} + \delta$, back to step 5;     \% Avoid infinite loops
          }
          \ENDIF
          \ENDIF
    \IF {$\lambda >= \varepsilon_{\lambda}$ and $|f(\theta^{k+1})| > \varepsilon_2$}
    \STATE $\lambda = \frac{\lambda}{2}$, back to step 5;     \% Newton downhill iteration
    \ENDIF
    }
    \ENDIF
    \end{algorithmic}
\end{algorithm}

\subsection{Lyapunov stability analysis of power grids}\label{Lyapunov stability}
Roughly, a power system is inevitable to suffer external small disturbances, and an equilibrium is considering stable if the system can restore to the original equilibrium with sufficient accuracy after small disturbances disappear (Fig.~\ref{smallnetwork}\textbf{d-e}). If the system moves away from the equilibrium after small disturbances, then the equilibrium is unstable (Fig.~\ref{smallnetwork}\textbf{f}). In this subsection, we will definite the Lyapunov stability of the power system and subsequently derive the stability criteria.

The mathematical definition of Lyapunov stability of power systems is: $\forall$ $\varepsilon > 0$, $\exists$ $\delta(\varepsilon, t_0)$, if any point $\boldsymbol{\theta}(t)$ on the trajectory staring from any initial state in the set $\{\boldsymbol{\theta^0}: \|\boldsymbol{\theta^0} - \boldsymbol{\theta^e} \| < \delta(\varepsilon, t_0)\}$ satisfies: $\| f(\boldsymbol{\theta(t)}; \boldsymbol{\theta^0}, t_0)\| \leq \varepsilon, t_0 \leq t \leq \infty$, the system is stable (Fig.~\ref{smallnetwork}\textbf{d}). If $\lim_{t \rightarrow \infty}\|\boldsymbol{\theta(t)} - \boldsymbol{\theta^e} \| = 0$, the system is asymptotically stable (Fig.~\ref{smallnetwork}\textbf{e}). Otherwise, the system is unstable (Fig.~\ref{smallnetwork}\textbf{f}). From the definition, we can summarize that the Lyapunov stability requires the trajectory of the stable system from any point near the equilibria always stays a certain range of the equilibria rather than restores to the original equilibria. Therefore, it is more extensive than the traditional stability.

Then, we will introduce the Lyapunov stability criterion of power system at equilibrium $\boldsymbol{\theta^e}$ and the potential function for $\boldsymbol{\theta}$ is defined as:
\begin{equation}\label{vx}
  V(\boldsymbol{\theta}) = -\sum_{i = 1}^{N} P_i\theta_i - \frac{1}{2} \sum_{i,j = 1}^{N} K_{ij}\cos(\theta_i - \theta_j)
\end{equation}
From Eq.~(\ref{equilibria}), $\frac{\partial V}{\partial \theta_i}|\theta_i = \theta_i^e = 0, \forall i$ and thus $\boldsymbol{\theta^e}$ is a extremum of $V(\boldsymbol{\theta})$. The Hesse matrix $H$ of the potential function $V$ is:
\begin{equation}\label{DF}
 DF = \frac{\partial f}{\partial \boldsymbol{\theta}} |_{\boldsymbol{\theta}=\boldsymbol{\theta^e}}
   = \left(
  \begin{array}{c}
   \frac{\partial f_1}{\partial \theta_1} \quad \frac{\partial f_1}{\partial \theta_2} \quad \cdots \quad  \frac{\partial f_1}{\partial \theta_N} \\
   \\
   \frac{\partial f_2}{\partial \theta_1} \quad \frac{\partial f_2}{\partial \theta_2} \quad \cdots \quad  \frac{\partial f_2}{\partial \theta_N} \\
   \\
    \vdots \qquad \vdots \qquad \ddots \qquad \vdots \\
    \\
   \frac{\partial f_N}{\partial \theta_1} \quad \frac{\partial f_N}{\partial \theta_2} \quad \cdots \quad  \frac{\partial f_N}{\partial \theta_N} \\
  \end{array}\right)
\end{equation}
\begin{equation}\label{H}
  \begin{split}
 &DF_{ij} =
  \left\{
  \begin{array}{c}
  \sum_{j \neq i} \frac{-\sigma_{ij}}{N} A_{ij} \cos(\theta_{j}^{e} - \theta_{i}^{e}) \qquad i = j \\
  \\
  \frac{\sigma_{ij}}{N} A_{ij} \cos(\theta_{j}^{e} - \theta_{i}^{e}) \qquad Otherwise
  \end{array}
  \right.  \\
 & H = -DF
\end{split}
\end{equation}
Based on the Hesse matrix $H$, the Lyapunov stability criterion of power systems is derived in Lemma~\ref{criterion}.
\begin{theorem}\label{criterion}
For a given system (network) $G$ and a equilibrium $\boldsymbol{\theta^e}$, $\boldsymbol{\lambda} = [\lambda_1, \lambda_2, \cdots, \lambda_N]$ are the eigenvalues of $H$ and are ranked by their real part such that $\lambda_1 = 0$~\cite{manik2017cycle} and $\lambda_2 < \lambda_3 < \cdots <\lambda_N$.
 \begin{itemize}
   \item If $Re(\lambda_2 > 0)$, the system is asymptotically stable (Fig.~\ref{smallnetwork}\textbf{c}). $\mathop {\lim }\limits_{t \to \infty } \left\| {\boldsymbol{\theta}(t) - \boldsymbol{{\theta^e}}} \right\| = 0$.
   \item If $Re(\lambda_2 = 0)$, the system $G$ is stable in the sense of Lyapunov (Fig.~\ref{smallnetwork}\textbf{d}). $\mathop {\lim }\limits_{t \to \infty } \left\| {\boldsymbol{\theta}(t) - \boldsymbol{\theta^e}} \right\| \leq \varepsilon$.
   \item If $Re(\lambda_2 < 0)$, the system $G$ is unstable (Fig.~\ref{smallnetwork}\textbf{f}). $\mathop {\lim }\limits_{t \to \infty } \left\| {\boldsymbol{\theta}(t) - \boldsymbol{\theta^e}} \right\| > \varepsilon$.
 \end{itemize}
\end{theorem}

\section{Dynamically induced cascading failures in power grids}
In this section, we focus on analyzing cascading failures after the network is attacked from a dynamic perspective. There are two types of failures: overload failures and unstable failures. The former mainly occurs in transmission lines and the latter always induce the malfunction of machines.

\subsection{Overload failures of lines}
\begin{figure}
  \centering
  \includegraphics[width=8.5cm]{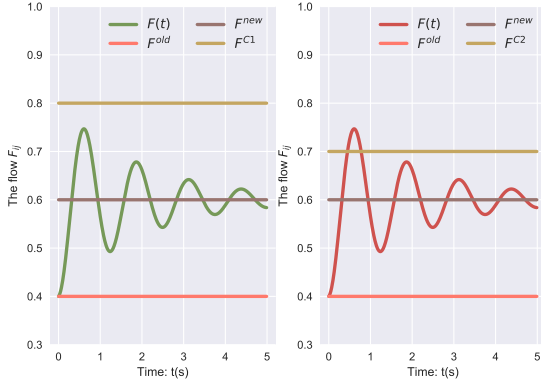}
  \caption{The conversion procession of flow in transmission line. $F^{old}$ is the flow of the original network and $F^{new}$ is the flow of the residual network after attack. $F^{C1}$ and $F^{C2}$ denote two different capacities of the transmission line. \textbf{a} The green solid line indicates $F_{ij}$ is smoothly converted after the network structure has changed. \textbf{b} The red solid line indicates the $E_{ij}$ will shutdown since the maximal flow $F^{max}$ exceeds the carrying capacity $C_{ij}$.}\label{over}
\end{figure}

The original network $G$ is usually assumed running stably at the equilibria $\boldsymbol{\theta^0}$ (N-0 stable). The steady flow of $G$ is denoted as $F^{old}$. After the network is attacked and subsequently causes line failure, Sc\"{a}hfer et al.~\cite{schafer2018dynamically} studied the case that the residual network $G'$ will stably operate at the new equilibria $\boldsymbol{\theta'}$ (N-1 stable) and the network flow is $F^{new}$. They found the conversion process of the steady flow ($F^{old} \rightsquigarrow F^{new}$) is not abrupt but the oscillations converge. Hence this process can be approximated by a damped sinusoidal function of time \cite{schafer2018dynamically}:
\begin{equation}\label{flowchange}
\begin{split}
    & F_{ij}(t) \approx F_{ij}^{new} - \Delta F_{ij}\cos(\nu_{ij}t)e^{-Dt} \\
    & \Delta F_{ij} = F_{ij}^{new} - F_{ij}^{old}
\end{split}
\end{equation}
The maximal flow in this process can be roughly approximated by Eq.~(\ref{fmax}).
\begin{equation}\label{fmax}
  F_{ij}^{max} \approx F_{ij} + 2\Delta F_{ij}
\end{equation}
The capacity $C_{ij}$ of line $E_{ij}$ is proportional to $K_{ij}$:
\begin{equation}\label{capacity}
  C_{ij} = \alpha K_{ij}
\end{equation}
where $\alpha$ ($0 < \alpha <1$) is the tolerance parameter of lines. Hence $E_{ij}$ will shutdown when the maximal flow $F_{ij}^{max}$ exceeds its carrying capacity $C_{ij}$. For example, the small grid in Fig.~\ref{smallnetwork} is attacked by removing $E_{23}$, and the conversion procession of flow is shown in Fig.~\ref{over}. In traditional static network flow analysis methods, the corresponding line is safe since those methods assume that the flow jumps from $F^{old}$ to $F^{new}$ and $F(t)$ is within the limits of $F^{C1}$ and $F^{C2}$. Nevertheless, from Sc\"{a}hfer et. al.'s theory, the conversion process is dynamic rather than static and whether the line will failure depends on its carrying capacity $C$ and its maximal flow $F^{max}$ during the dynamic oscillation process. Therefore, the damped sinusoidal process of flow can detect the overload line which is ignored by the static flow methods.

\subsection{Small disturbance induced failures of nodes}\label{second failure}
\begin{figure}
  \centering
  \includegraphics[width=8.5cm]{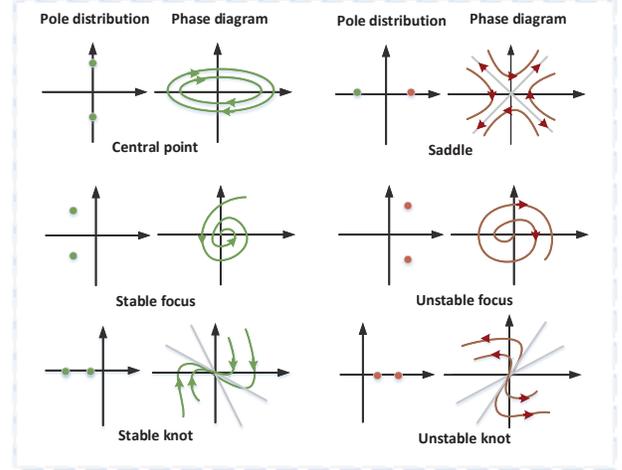}
  \caption{Phase trajectory maps of different poles distributions. In pole distribution maps, the horizontal axis is the real axis and the vertical axis is the imaginary axis. In phase diagram, the horizontal axis denotes $\theta_i$ and the vertical axis denotes $\dot{\theta_i}$. The green solid lines indicate the corresponding node is reliable and the red solid lines indicate the corresponding node is exceptional.}\label{phase}
\end{figure}

\begin{figure}
  \centering
  \includegraphics[width=8cm]{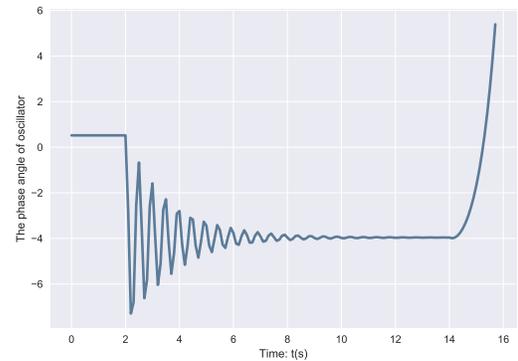}
  \caption{The state trajectory of $N_2$. In the original network $G$, the equilibrium of this node is $\theta^0_2 = 0.5186$. After the network is attacked, the node converges to $\theta'_2 = -3.9623$. However, the new equilibrium $\theta'$ is exceptional. At time $t = 14s$, a external small disturbance is applied to this node and it quickly diverges.}\label{node2}
\end{figure}

In this subsection, for the sake of universality, we abandon the assumption that the network is N-1 stable after the attack. For a given network, we use Lyapunov criterion (Lemma~\ref{criterion}) to judge the stability of the network. If the network is stable, each machine will smoothly operate at its equilibrium and the power system can generate steady flows to users. Conversely, the network will malfunction when it is unstable at the equilibrium. In this case, the Lyapunov criterion can only judge the stability of the network but fails to identify the nodes or lines that are invalid. In this paper, the small disturbance analysis method is presented to overcome the drawback of Lemma~\ref{criterion} and we find a new failure mode based on this method. The core idea of this method is that it is inevitable for power systems to suffer the external small disturbance $\Delta \boldsymbol{\theta} (\|\Delta \boldsymbol{\theta}\| < \varepsilon)$ and machine $i$ is considering reliable if it can return to the original equilibrium after small disturbance $\Delta \theta_i$. If the state trajectory $\theta(t)_i$ is divergent after applying the disturbance $\Delta \theta_i$, the machine (node) $i$ will be damaged and the lines connected with node $i$ will shutdown duo to overload (From Eq.~(\ref{flow}), the flow of $E_{ij}$ will exceed its capacity when $\theta_j$ is a constant but $\theta_i$ diverges). The detailed algebraic derivation is shown as follows:

Let $\boldsymbol{\theta^e} = [\theta_1^e, \theta_2^e, \cdots, \theta_N^e]$ be a equilibrium of unstable system $G$. From Eq.~(\ref{equilibria}), $\forall$$i$, $\theta_i^e$ satisfies:
\begin{equation}\label{se}
  P_i + \sum_{j = 1}^{N} K_{ij} \sin(\theta_j^e - \theta_i^e)
\end{equation}
Now, we apply a small disturbance $\Delta \boldsymbol{\theta}$ to network $G$:
\begin{equation}\label{AddDisturbance}
  \begin{split}
  & \boldsymbol{\theta} = \boldsymbol{\theta^e} + \Delta \boldsymbol{\theta} \Rightarrow \theta_i = \theta^e_i + \Delta\theta_i \\
  & \dot{\boldsymbol{\theta}} = \Delta \dot{\boldsymbol{\theta}} \Rightarrow \dot{\theta_i} = \Delta \dot{\theta_i}\\
  & \ddot{\boldsymbol{\theta}} = \Delta \ddot{\boldsymbol{\theta}} \Rightarrow \ddot{\theta_i} = \Delta \ddot{\theta_i}
  \end{split}
\end{equation}
Substitute Eq.~(\ref{AddDisturbance}) into Eq.~(\ref{swing eqaution}):
\begin{equation}\label{DisturbanceSW}
 M_i \Delta \ddot{\theta_i} + D_i \Delta \dot{\theta_i} = P_i + \sum_{j = 1}^{N} K_{ij} \sin(\theta_j^e - \theta_i^e + \Delta \theta_j - \Delta \theta_i)
\end{equation}
The $\sin$ term of Eq.~(\ref{DisturbanceSW}) can be expanded by Taylor formula:
\begin{equation}\label{Taylor}
  \begin{split}
      & \sin(\theta_j^e - \theta_i^e + \Delta \theta_j - \Delta \theta_i) \\
      & = \sin(\theta_j^e - \theta_i^e) + \cos(\theta_j^e - \theta_i^e)(\Delta \theta_j - \Delta \theta_i) + O(\Delta \theta_j - \Delta \theta_i)^2 \\
      & \approx \sin(\theta_j^e - \theta_i^e) + \cos(\theta_j^e - \theta_i^e)(\Delta \theta_j - \Delta \theta_i)
  \end{split}
\end{equation}
The reason we abandon the term of $O(\Delta \theta_j - \Delta \theta_i)^2$ is that the norm of the small disturbance $\Delta \boldsymbol{\theta}$ is quite small. Then, by combining Eq.~(\ref{se}) and Eq.~(\ref{Taylor}), we can deduce:
\begin{equation}\label{Psin}
\begin{split}
  & P_i + \sum_{j = 1}^{N} K_{ij} \sin(\theta_j^e - \theta_i^e + \Delta \theta_j - \Delta \theta_i)  \\
  & = P_i + \sum_{j = 1}^{N} K_{ij} \sin(\theta_j^e - \theta_i^e) + \cos(\theta_j^e - \theta_i^e)(\Delta \theta_j - \Delta \theta_i) \\
  & = P_i + \sum_{j = 1}^{N} K_{ij} \sin(\theta_j^e - \theta_i^e) + \sum_{j = 1}^{N} K_{ij} \cos(\theta_j^e - \theta_i^e)(\Delta \theta_j - \Delta \theta_i) \\
  & = \sum_{j = 1}^{N} K_{ij} \cos(\theta_j^e - \theta_i^e)(\Delta \theta_j - \Delta \theta_i)
\end{split}
\end{equation}
Substitute Eq.~(\ref{Psin}) to Eq.~(\ref{DisturbanceSW}):
\begin{equation}\label{simpleSW}
  M_i \Delta \ddot{\theta_i} + D_i \Delta \dot{\theta_i} = \sum_{j = 1}^{N} K_{ij} \cos(\theta_j^e - \theta_i^e)(\Delta \theta_j - \Delta \theta_i)
\end{equation}
Let $\beta_{ij} = K_{ij} \cos(\theta_j^e - \theta_i^e)$ and $\sum_{j}\beta_{ij} = \beta_i$. Then:
\begin{equation}\label{finalSW}
 M_i \Delta \ddot{\theta_i} + D_i \Delta \dot{\theta_i} =  \sum_{j}\beta_{ij} \Delta \theta_j - \beta_i \Delta \theta_i
\end{equation}
The characteristic equation of Eq.~(\ref{finalSW}) is:
\begin{equation}\label{ce}
  M_i s^2 + D_i s + \beta_i = 0
\end{equation}
The roots (poles) of this equation are:
\begin{equation}\label{roots}
  \begin{split}
      & s_1 = \frac{-D_i + \sqrt{D_i^2 - 4M_i\beta_i}}{2M_i} \\
      & s_2 = \frac{-D_i - \sqrt{D_i^2 - 4M_i\beta_i}}{2M_i}
  \end{split}
\end{equation}
Therefore, the solution of Eq.~(\ref{finalSW}) is:
\begin{itemize}
  \item If $s_1 = s_2; s_1,s_2 \in \mathbb{R}$, then $\Delta \theta_i(t) = (C_1 + C_2t)e^{s_2t}$
  \item If $s_1 \neq s_2; s_1,s_2 \in \mathbb{R}$, then $\Delta \theta_i(t) = C_1e^{s_1t} + C_2e^{s_2t}$
  \item If $s_1 = \mu + \nu i, s_2 = \mu - \nu i; s_1, s_2 \in \mathbb{C}$, $\Delta \theta_i(t) = e^{\mu t} (C_1\cos(\nu t) + C_2\sin(\nu t))$
\end{itemize}
where $C_1$ and $C_2$ are constants. According to the distribution of poles, the corresponding phase diagrams are shown in Fig.~\ref{phase}. Based on the solution of Eq.~(\ref{finalSW}) and Fig.~\ref{phase}, we can derive the following Lemma:
\begin{theorem}\label{exception}
Let $\boldsymbol{\theta} = [\theta_1, \theta_2, \dots, \theta_N]$ be one equilibrium of the network $G$ and $\Delta \boldsymbol{\theta} = [\Delta \theta_1, \Delta \theta_2, \dots ,\Delta_N]$ be the small disturbance applied to $G$. If $\lim_{t \rightarrow \infty} \Delta \theta(t)_i = 0$, $N_i$ will stably operate at $\theta_i$. Conversely, if $\lim_{t \rightarrow \infty} \Delta \theta(t)_i = \infty$, $N_i$ is exceptional and the divergent state trajectory will induce the failure of $N_i$ and $E_{i:}$ (Eq.~(\ref{flow})). Let $f_i$ be the state equation of $N_i$ and $S_i$ be the characteristic equation of $f_i$. $s_1$ and $s_2$ are the roots of $S_i$. The criterion to determine whether $N_i$ is exceptional is:
\begin{itemize}
  \item If $Re(s_1) < 0$ and $Re(s_2) < 0$, then $\lim_{t \rightarrow \infty} \Delta \theta(t)_i = 0$, $\lim_{t \rightarrow \infty} \theta(t)_i = \theta_i$, the corresponding node is reliable.
  \item If $Re(s_1) = 0$ and $Re(s_2) = 0$, then $\lim_{t \rightarrow \infty} \Delta \theta(t)_i = c$, $\lim_{t \rightarrow \infty} \theta(t)_i = c + \theta_i$, the corresponding node is also reliable from the perspective of Lyapunov stability (Subsection~\ref{Lyapunov stability}).
  \item If $Re(s_1) > 0$ or $Re(s_2) > 0$, then $\lim_{t \rightarrow \infty} \Delta \theta(t)_i = \infty$, $\lim_{t \rightarrow \infty} \theta(t)_i = \infty$, the corresponding node is exceptional.
\end{itemize}
\end{theorem}
From Eq.~(\ref{roots}), we can deduce that the convergence or divergence rate of $\Delta \theta_i$ determined by $D_i, M_i, \beta_i$. Here, we use an instance (Fig.~\ref{smallnetwork}\textbf{a}) to illustrate the existence of this failure mode. The parameters we used in this small network are shown in Table~\ref{smallresult}. Originally, the network $G$ was stably running at the equilibrium $\boldsymbol{\theta^0}$. Then, we attacked network $G$ by erasing the line $E_{23}$ and the new equilibrium became $\boldsymbol{\theta'}$. Nevertheless, the new equilibrium was unstable judged by Lemma~\ref{criterion}. In particular, the dynamic behavior of $N_2$ was exceptional after a small disturbance observed by Eq.~(\ref{finalSW}), Eq.~(\ref{roots}), Fig.~\ref{phase}, and Lemma~\ref{exception}. The state trajectory of $N_2$ is shown in Fig.~\ref{node2}. The curve first oscillates to the new equilibria $\theta'_2$ during $t = [2, 12]$. At $t = 14$, $N_2$ is disturbed. The state trajectory of $N_2$ quickly diverges and then induces the failure of $N_2$ and $E_{2:}$ (the blue edges of Fig.~\ref{smallnetwork}\textbf{a}). Further, the cascade mechanism is triggered and thereupon induces the failure of $N_4$ and $N_1$. Finally, the network will collapse with cascade propagation.

\begin{table}
\footnotesize
\caption{The simulation results of the small network}\label{smallresult}
\setlength{\tabcolsep}{1mm}{
\begin{tabular*}{\columnwidth}{@{\extracolsep{\fill}}@{~~}llcccc@{~~}}
\toprule
 Node   &Power                       &Equilibrium              &Equilibrium       &$s_1$         &$s_2$         \\
 ID             &                    &Before attack            &After attack                                  \\
\midrule
(1)               &-1                  & 0.2453                  & 2.2791         &$-0.3 + 0.20i$ & $-0.3 - 1.20i$  \\
(2)	              &1.5                 & 0.5186                  & -3.9632        &$1.43$        & $-2.03$  \\
(3) 	          &-1                  & 0.1284                  & 1.9240         &$-0.3 + 1.74i$ & $-0.3 - 1.74i$  \\
(4)   	          &-1                  & 0.2453                  & -4.4794        &$-0.3 + 1.21i$ & $-0.3 - 1.21i$  \\
(5) 	          &1.5                 & 0.7234                  & 3.4478         &$-0.3 + 1.68i$ & $-0.3 - 1.68i$  \\
\bottomrule
\end{tabular*}}
\end{table}

\begin{figure}
  \centering
  \includegraphics[width=8.5cm]{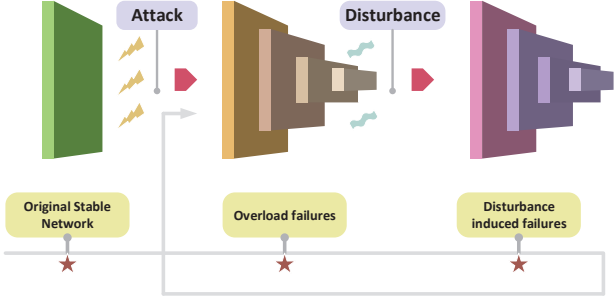}
  \caption{Dynamically induced cascading failures of power grids. An attack on the initial stable network triggers cascading overload failures of lines and may lead to two results: the network crashes or rebalances. In the second case, the new equilibrium may unstable and some nodes will fail duo to the small disturbance. The residual network will then form a new equilibrium again. This process will loop forever until the stability condition is met, or the network is completely paralyzed.}\label{cascade}
\end{figure}

Integrated the failure modes described in this section, the complete cascading failures process in power grids is shown in Fig.~\ref{cascade}. For the sake of simplicity, we retain the null hypothesis that the initial network is stable but the presented method is still applicable to the unstable initial network.

\section{Results} \label{attack network}
Above we roughly used a small network to illustrate the effectiveness of the presented method. In this section, the Spanish grid is introduced to verify the applicability of our method in a real scenario. Duo to the overload failure mode has been well analyzed in Sc\"{a}hfer et al.'s work~\cite{schafer2018dynamically}, we mainly focus on cascading failures induced by the small disturbance in our experiment. In this case, we avoid the overload failures by increasing the tolerance $\alpha$ in Eq.~(\ref{capacity}) and thereby the malfunction of nodes and lines is purely caused by the small disturbance. The standard procedures for our experiment are shown in Algorithm~\ref{experiment}.
\begin{algorithm}
\caption{Small disturbance induced cascading failures}\label{experiment}
\textbf{Input:} Network $G$; Number of attacks $T$. \\
\textbf{Initial:} The number of failed nodes $FN = 0, FN \in \mathbb{R}^{T}$ and failed edges $FE = 0, FE \in \mathbb{R}^{T}.$\\
\textbf{Output:} $J \in \mathbb{R}^{N \times T}, FN, FE.$
 \begin{algorithmic}[1]
    \FOR {$i = 1$ to $T$}
    \STATE $G' = G$;
    \STATE Calculate the equilibrium $\boldsymbol{\theta}$ of $G'$(Algorithm~\ref{Newton});
    \STATE Determine whether $G'$ is stable using Lemma~\ref{criterion}. If not, back to step 8, else, \textbf{continue};
    \STATE Orderly or randomly remove an edge from $G'$; \% attack
    \STATE Calculate the new equilibrium $\boldsymbol{\theta'}$ of the residual network by Algorithm~\ref{Newton}. Calculate the overload lines in the conversion process: $\boldsymbol{\theta} \rightarrow \boldsymbol{\theta'}$ based on Eqs.~(\ref{flowchange}-\ref{capacity});
    \STATE Determine whether $G'$ is stable using Lemma~\ref{criterion}. If not, \textbf{continue}, else, \textbf{break};
    \STATE Determine whether $N_j, j \in [1, N]$ is reliable using Lemma~\ref{exception}. If exceptional, delete $N_{j}$, $J_{ij} = 1$, else, $J_{ij} = 0$, $\boldsymbol{\theta'} = \boldsymbol{\theta}$;
    \STATE Count the number of failed nodes $FN[i]$ and edges $FE[i]$ in secondary outage;
    \STATE Determine whether $G'$ paralyzes. If $G'$ paralyzes, \textbf{break}, else, back to step 6;
    \ENDFOR
    \end{algorithmic}
\end{algorithm}

\begin{figure}
  \centering
  \includegraphics[width=8.5cm]{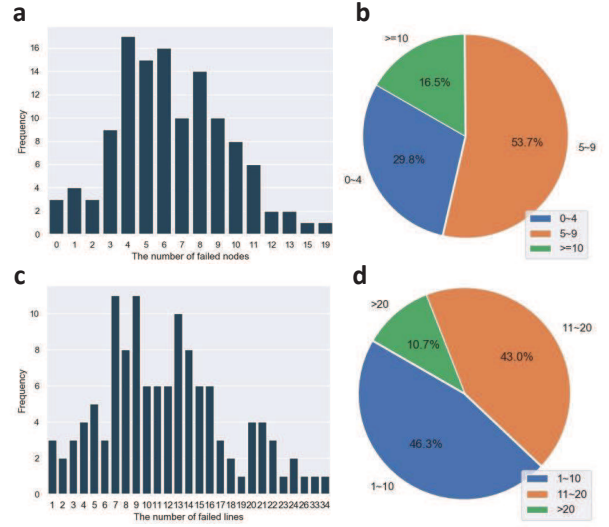}
  \caption{The distribution of secondary outages in Spanish grid. In this experiment, the inertia term $M_i = 1$, the coupling strength $K = 8$, the damping coefficient $D =0.5$(the same as below). The histograms show the number of nodes or edges that failed after each attack in 121 attacks. The pies show the probability of the failed number of nodes or edges after an attack.}\label{failure}
\end{figure}

In our experiment, the initial network $G$ stably operates at its equilibrium $\boldsymbol{\theta}$ ( Algorithm~\ref{experiment} can also simulate the case that the initial network is unstable). In fact, the phase difference $|\theta_j - \theta_i|, \forall i,j$ is quite small in most synchronous states. Even if the network is attacked, the network can remain in a new stable equilibrium with a high probability. That implies cascading failures induced by small disturbances are difficult to occur. In our experiment, to observe this kind of failure mode, we have made the network more vulnerable by appropriately adjusting the model parameters (e.g. $P_i$) to increase the phase difference between different oscillators. we attack network $G$ 350 times by orderly removing one line in each attack, of which 121 successfully triggered cascading failures. The macro data distributions of secondary outages are shown in Fig.~\ref{failure}. On average, about $10\%$ of the nodes failed duo to external small disturbances in secondary outages. Therefore, small disturbances can cause the local malfunction of the unstable network in secondary outages and cascading propagation will further lead to the paralysis of the entire network.  In summary, cascading failures triggered by small disturbances are extremely difficult to occur but can cause power outages throughout the grid.

\begin{figure}
  \centering
  \includegraphics[width=8cm]{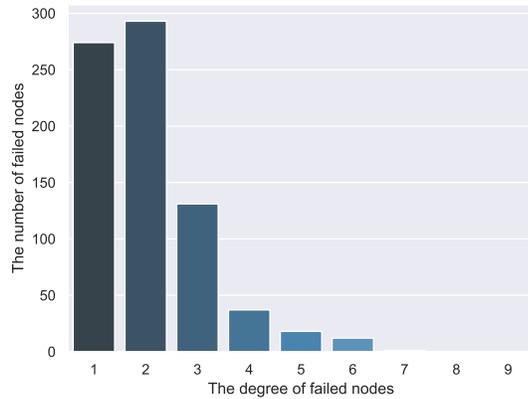}
  \caption{The degree distribution of the failed nodes in secondary outages. This histogram counts the degree of failed nodes in 121 attacks.}\label{degree}
\end{figure}

\begin{figure*}
  \centering
 \subfloat[$D_i = 0.5$]{
    \label{N14a} 
    \includegraphics[width=8cm]{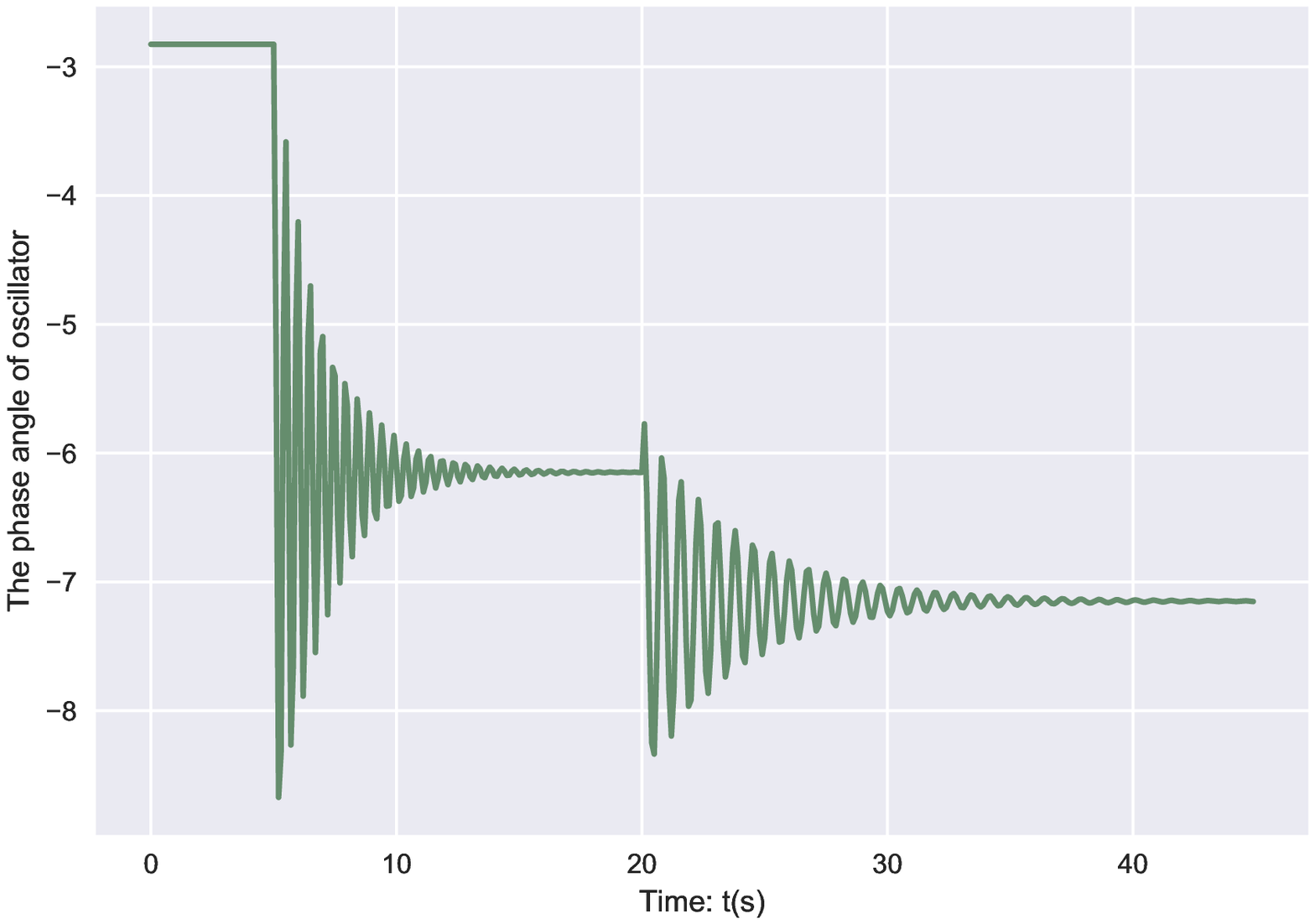}}\
  \subfloat[$D_i = 1$]{
    \label{N14b} 
    \includegraphics[width=8cm]{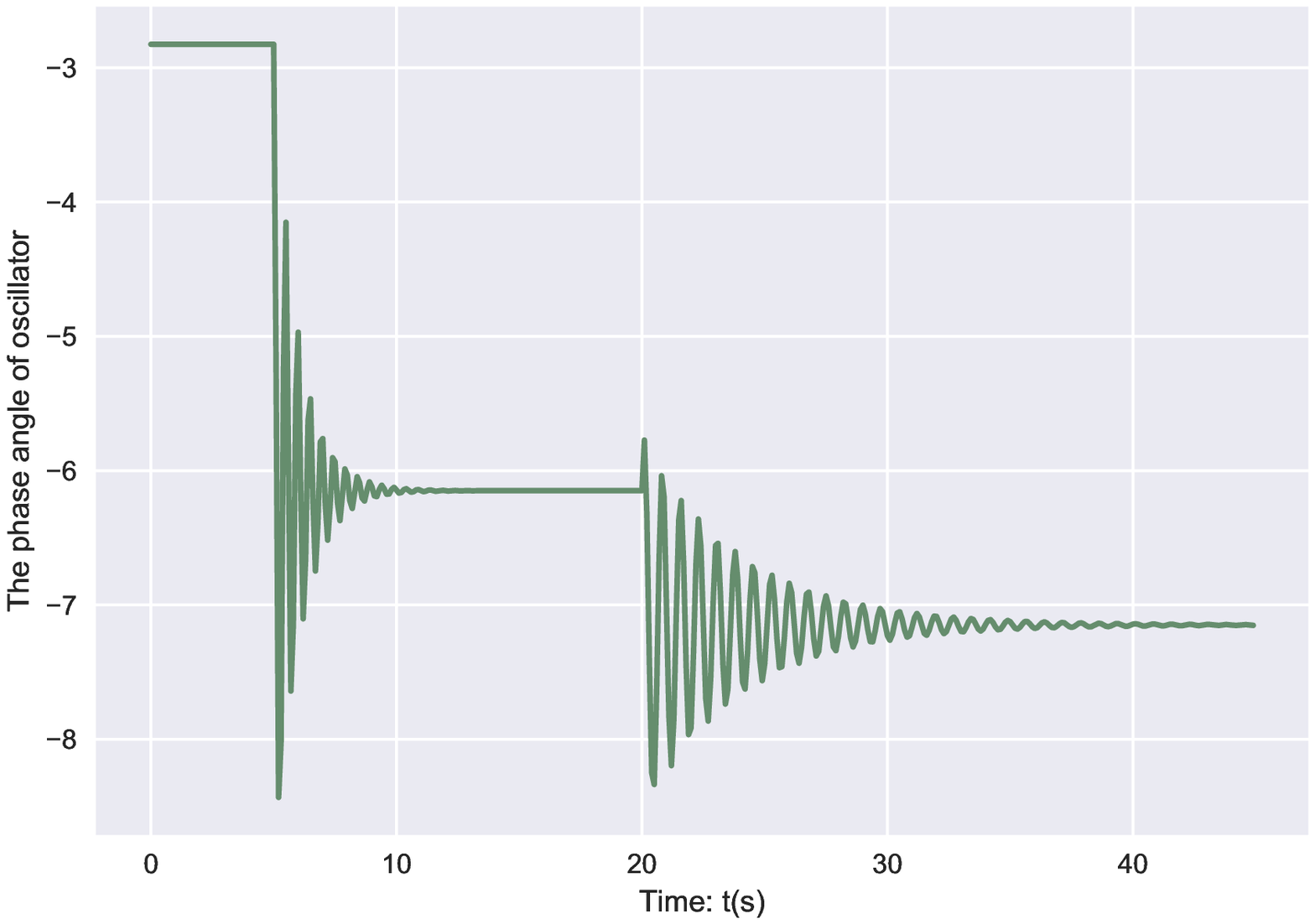}}\\
  \caption{The phase trajectory of $N_{14}$ under different damping coefficients $D_i$. After the original network is attacked, the residual network tends to a new equilibrium that is stable. From \textbf{a} and \textbf{b}, the convergence rate depends on the damping coefficient. At $t = 20s$, the system is disturbed and the phase trajectory reconverges.}
  \label{N14} 
\end{figure*}

\begin{figure}
  \centering
  \includegraphics[width=8cm]{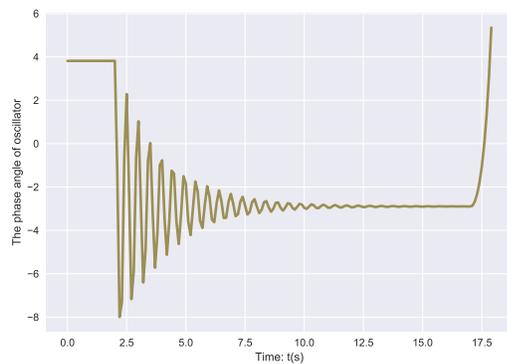}
  \caption{The phase trajectory of exceptional $N_{10}$. $N_{10}$ is also unreliable judged by Lemma~\ref{exception}. After being disturbed, the phase trajectory will diverge exponentially.}\label{N10}
\end{figure}

In Fig.~\ref{degree}, we have counted the degree of failed nodes in secondary outages. Faults tend to occur first in less-degree nodes and then indirectly cause the failure of hub nodes through cascade propagation. It is worth noting that the speed of cascade propagation is extremely fast and we will explain this phenomenon from a microscopic perspective. We have selected two typical nodes ($N_{14}: d(14) = 9$ and $N_{10}: d(10) = 1$) to observe their phase trajectory (Figs.~\ref{N14}-\ref{N10}) in Spanish grid. $N_{14}$ is reliable and it can return to a new equilibrium after being attacked or disturbed. The rate of convergence depends mainly on its damping coefficient $D_{14}$. $N_{10}$ is exceptional and its phase trajectory will divergence exponentially after being disturbed and further lead to the failure of itself and its neighbors. From Fig.~\ref{N10} ($t>17s$), the propagation of faults along adjacent nodes is particularly fast, usually less than $0.5s$.

From the above data analysis, we can summarize some characteristics of cascading failures caused by small disturbances:
\begin{enumerate}
  \item Cascading failures induced by small disturbances rarely happen but can have devastating consequences.
  \item Cascade propagation is so fast that it can easily escape the grid protection mechanism.
  \item The propagation path of this kind of failure mode is from edge nodes to central nodes.
\end{enumerate}
These characteristics are clearly reflected in the large-scale blackouts in recent years.

\section{Discussion}\label{discussion}
In this paper, motivated by the advance of dynamically induced cascading failures in power grids, we proposed a more general model to analyze cascading failures and its spread mechanism. Our model abandoned the assumption existing in the traditional failure models that the network is stable. Instead, the Lyapunov criterion was introduced to judge the stability of the network. Based on this theory, besides the overload failures of lines during the steady flow conversion process, we found that small disturbances can also induce the shutdown of exceptional nodes and further trigger cascade in unstable networks. Through data analysis, this kind of failure mode has the following characteristics: low incidence, large destructiveness, and fast propagation speed. Edge (leaf) nodes are first affected and then lead to the failure of a large fraction of the transmission grid. The failure will propagate to the central nodes with an extremely fast rate and result in the paralysis of the entire network. These characters are consistent with the blackouts observed in real power grids.

Although we tried to construct a comprehensive dynamic model to study cascading failures and achieved the desired results, there are still many challenges before the large-scale power outage of the grid is fully uncovered. For instance, how to reasonably allocate the power $P$ of the failed nodes to their neighbors to bring the network to a new equilibrium. This is also the reason that we just calculate the data of secondary outages rather than the final outages in our experiment. There may be no equilibrium in the network because the inappropriate $P$ may result in no solution to the swing equations (Eq.~(\ref{equilibria})). Additionally, how to force the exceptional node to the stable state before cascade propagation is also a ticklish problem. The usual solution is to design a controller to stabilize the exceptional nodes. Nonetheless, such controllers are always of low-efficiency duo to the propagation speed is extremely fast (Fig.~\ref{N10}). Meanwhile, when we apply control to the exceptional nodes, it will inevitably affect the equilibrium of their neighbors. How to decouple the complex system presents challenges in designing controllers. In our future work, we will aim to solve the above problems to ensure the safe and stable operation of power grids.


%

%



\ifCLASSOPTIONcaptionsoff
  \newpage
\fi



\bibliographystyle{IEEEtran}
\bibliography{IEEEabrv,IEEEexample}
\end{document}